\documentstyle[12pt,cite]{article}                                        
\catcode`\@=11                                                       
\def\ee{\end{equation}}
\def\be{\begin{equation}}
\def\la{\label}

\def\N{{\tt N}}
\def\1{{\bf 1}_N}

\begin{document}                                                     

\renewcommand{\theequation}{\thesection.\arabic{equation}}        
\newcommand{\mysection}[1]{\setcounter{equation}{0}\section{#1}}  

\begin{titlepage}


\hfill{SPIN -- 2001/18}

\vspace{2 cm}

\centerline{{\huge{On the Supersymmetric Index}}}
\centerline{{\huge{of the M-theory 5-brane}}} 
\centerline{{\huge{and Little String Theory}}} 

\vspace{2 cm}

\centerline{\large{Giulio Bonelli}}
\centerline{Spinoza Institute, University of Utrecht}
\centerline{Leuvenlaan 4, 3584, CE Utrecht, The Netherlands}
\centerline{G.Bonelli@phys.uu.nl}

\vspace{4 cm}

{\it Abstract}: 
We propose a six-dimensional framework
to calculate the supersymmetric index of M-theory 5-branes wrapped
on a six-manifold with product topology $M_4\times T^2$,
where $M_4$ is a holomorphic 4-cycle in a Calabi-Yau three-fold.
This is obtained by zero-modes counting of the self-dual tensor contribution plus
``little'' string states and correctly reproduces
the known results which can be obtained 
by shrinking or blowing the $T^2$ volume parameter.
We also extract the geometric moduli space of the multi M5-brane system
and infer the generic structure of 
the supersymmetric index for more general geometries.

\end{titlepage}

\section{Introduction}

A proper definition of M-theory as a non perturbative 
framework for superstring theories is still 
an open problem. It is strongly shaded among the others
by the lack of understanding the very structure of the 
world-volume theory of the M5-branes.

As far as this sector of the theory is concerned, its totally decoupled
phase has been named little string theory
\cite{bps5,DVVx, Mm}. Little string theory (for a review, see also \cite{ofer,pa}
and references therein) 
is still anyway poorly understood.
Its low energy limit is accepted to be a theory of
$(0,2)$ self-dual tensor multiplets which lacks both locality and a Lagrangean 
formulation and can be therefore studied by now just with constructive 
methods. The UV complete theory has been conjectured \cite{bps5}
to be a closed string theory in 6 dimensions describing
the boundary states of the membranes ending on the 5-branes.
This strong characterization of M-theory has passed several consistency checks
\cite{bps5,5dbh,Mm,ofer}.
Specifically, little string theory has been extensively studied on tori
and orbifold K3, while much less is known about it in more general cases.
Anyway it turns out \cite{agmh} that the appropriate nature of the degrees of freedom 
which complete in the UV the interacting theory is stringy and this further
enforces the above proposal.

The picture that is coming out and that we have in mind is the following.
Suppose one is dealing with a bounce of M5-branes. In the low energy approximation,
if they can be separated neatly one from each other, each brane hosts 
a self-dual tensor multiplet theory describing its effective degrees
of freedom. They come as zero-modes of membranes ending on them.
Suppose now that we take two M5-branes close-by.
In this case, at some distance, the effective theory is expected to develop
interacting terms corresponding to an analog of gauge symmetry enanchement
$U(1)\times U(1)\,\to\, U(2)$, but it turns out that 
these kind of structure -- i.e. a higher rank generalization of non-abelian gauge symmetry --
does not exists in a local sense.
Notice that this fact is natural because the boundary of the membranes stretched 
between the M5-branes are stringy objects while a local field theory
interaction would describe particle vertices.
This situation has to be compared with the other situation in which D-branes come together:
in this case the boundary of the strings stretched between the D-branes is composed of point-like
objects whose field theory description turns out to be given by the non
abelian gauge degrees of freedom via the usual Chan-Paton construction.
The outcome of this analysis (see the cited literature for more complete
treatment) is that the effective interacting theory of M5-branes has to be described as 
a string theory in six dimensions. The aim of this paper is to check this 
picture with a supersymmetric index calculation.
This means that we will start from a known result for this object (in a particular
geometrical set-up) which has been already obtained by other methods
and we will recalculate it from a six-dimensional string theory point of view.
This will be done within an on-shell model. The off-shell formulation of 
the six dimensional string theory is out of reach in the present paper
and an open problem.

In this paper, in fact, we use the analysis performed in \cite{5b} by proposing 
a six dimensional framework for the BPS state counting of the
M5-brane which correctly
reproduces the results for the supersymmetric index
which have been obtained there in a dimensionally reduced framework.
This interpretation gives also some hints on the structure of the moduli space 
of supersymmetry preserving solutions of the M5-brane world volume theory which we check 
to be compatible with non-perturbative superstring dualities results.
In the framework that we propose, the complete counting of these configurations
amounts of two sectors which are the set of fluxes of the self-dual tensor
multiplet plus certain little string BPS saturated configurations.

This paper is organized as follows.
In the next section we will briefly review the results obtained in \cite{5b}
for the supersymmetric index of certain M5-branes configurations
in a specific geometrical setup.
In the subsequent section we will explore a six-dimensional point of view 
about the M5-brane supersymmetric index which correctly reproduces 
the previous results by a combined argument coming from the analysis
of the self-dual 2-form potential as counted in \cite{witten5,mans}
and the little strings to which we referred above.
In the next section we discuss some points about the multi 5-brane
result,
we completely determine the structure of the geometric moduli space
of the multi five-brane bound states
and infer the generic structure of the supersymmetric index for more general geometries.
This is done by combining together the informations encoded in the supersymmetric index
formula, superstring dualities and results about the structure of $(0,2)$ theories.
As an important by-product we check the apparence of extra massless string states 
corresponding to the interaction between M5-branes as they approach each others.
A final section is dedicated to the discussion about open questions.

\section{The M5-branes index on $T^2\times M_4$}

The geometric set-up that we refer to is the following \cite{5b,msw}.
We consider M-theory on $W=Y_6\times T^2\times R^3$, where
$Y_6$ is a Calabi-Yau threefold of general holonomy.
Let $M_4$ be a supersymmetric simply connected four-cycle in $Y_6$
which we take to be a representative of a very ample divisor.
Notice that $M_4$ is automatically equipped with a Kaehler form
$\omega$ induced from $Y_6$ and is simply connected.
We consider then $N$ M5-branes wrapped around $C=T^2\times M_4$.

Very few is known about the full world-volume theory describing a bunch
of parallel M5-branes, but from what we believe to be true in M-theory, 
we can extract already some informations about it.
It can be shown that in this specific geometrical set-up \cite{5b}
the potential anomalies which tend to ruin gauge invariance
of the world-volume theory are absent and that
it is then meaningful to define a supersymmetric index for the above 5-branes
bound states by extending the approach in \cite{vafa'}.
As it is well known, the supersymmetric index is independent on 
smooth continue parameters and
as a consequence we have that this counting of supersymmetry preserving 
states have to coincide in the large and small $T^2$ volume.
In \cite{5b} this calculation was performed and this equivalence was shown to be effective.
In particular, in the large $T^2$ volume we calculated the supersymmetric index
of the relevant two dimensional $\sigma$-model with target the moduli space
of susy-preserving configurations
of the corresponding twisted ${\cal N}=4$ SYM theory on $M_4$.
This was shown to consist of the Hilbert scheme of holomorphic coverings of 
$M_4$ in $Y_6$. 
On each stratum, characterized by the total rank of the covering $N$ and by
the topological numbers of 
the associated spectral surface $\Sigma$, we calculate the supersymmetric index as
\be
{\cal E}=
\frac{(Im\tau)^{d/2}}{V_d} {\rm Tr}_{RR}
\left[
(-1)^FF_R^{\sigma/2}q^{L_0}\bar q^{\bar L_0}
\right]\, ,
\label{si}\ee
where $d$ is the number of non-compact scalar bosons, $V_d$ their zero-mode volume and 
$(0,\sigma)$
are the supersymmetries of the model. By general arguments, ${\cal E}$ is a 
$(-d/2,-d/2)+(0,\sigma/2)$ modular form.

On the other side of the equivalence,
our explicit calculation, done by making use of the lifting technique
\cite{bbtt,MST} (see also \cite{tm}), for the 4 dimensional gauge theory gave, 
for the generic irreducible sector relative to the partition $N=\sum_a n_a\cdot a$
and to a given irreducible holomorphic covering of $M_4$ in $Y_6$ of rank $a$,
\be
{\cal E}_{n_a,a}= H_{n_a}
\sum_{\varepsilon}\frac{\theta_{\Lambda^{\Sigma_a}+x}}{\eta^{\chi_{\Sigma_a}}}
\label{eg}\ee
where 
$H_n$ is the Hecke operator of order $n$, $\varepsilon$ is a label for 
the square-roots of the canonical line bundle 
(spin structures) on $M_4$ 
with respect to a given one
as ${\cal O}_\varepsilon\otimes K^{1/2}$ with ${\cal O}_\varepsilon^2=1$,
$x=[{\cal O}_\varepsilon^{\otimes a+1}]$ shifts correspondingly the
lattice of integer periods $\Lambda^{\Sigma_a}$ on $H^2(\Sigma_a,R)$. 
The $\theta$-function on the lattice $\Lambda$ is defined as
\be
\theta_\Lambda(q,\bar q)=
\sum_{m\in\Lambda}q^{\frac{1}{4}(m,*m-m)}{\bar q}^{\frac{1}{4}(m,*m+m)}
\label{theta}\ee
and is a modular form of weight $(b_-/2,b_+/2)$.
$\eta(q)$ is the Dedekind $\eta$-function and $\chi=2+b_2^{\Sigma_a}$
is the Euler number of the spectral surface $\Sigma_a$.
In particular (\ref{eg}) for the single 5-brane reads
\be
{\cal E}=\frac{\theta_\Lambda(q,\bar q)}{\eta^\chi}
\la{single}\ee
which is a modular form of weight $(b_-/2,b_+/2)+(-\chi/2,0)=
(-1-b_+/2,b_+/2)$. 
All this agrees with the result obtained by the explicit evaluation of the 
supersymmetric index (\ref{si}). In particular, we have $\sigma=2b_++2=4b^{(2,0)}+4$
right fermions and $d=3+2b^{(2,0)}=2+b_+$ non-compact scalar bosons
by dimensional reduction giving ${\cal E}$ to be a modular form 
of total weight $(-1-b_+/2,b_+/2)$ as we just obtained.

\section{The single M5-brane case}

\subsection{The low energy contribution}

Let us start with the single M5-brane case.
The bosonic spectrum of the low energy world-volume theory of this 5-brane is given by
a 2-form $V$ with self-dual curvature and five real bosons taking values
in the normal bundle $N_C$ induced by the structure of the embedding
as $T_W|_C=T_C\oplus N_C$. Passing to the holomorphic part and to 
the determinants and using the properties of $Y_6$,
it follows that the five transverse bosons are respectively, three non-compact 
real scalars $\phi_i$
and one complex section $\Phi$ of $K_{M_4}=\Lambda^{-2} T^{(1,0)}_{M_4}$,
which is the canonical line bundle of $M_4$.

The (partially) twisted chiral $(0,2)$ supersymmetry completes the spectrum.
It is given by a doublet of complex 
anti-commuting fields 
which are $(2,0)-$forms in six dimensions 
and a doublet of complex anti-commuting fields
which are scalars in six dimensions. Notice that these fermionic content
reduces to the two relevant fermionic spectra in the large and small $T^2$ volume
limits respectively.

In principle there would be two multiplicative contributions to the supersymmetric index. 
One given by zero modes counting and another given by the one loop determinants.
In our case, anyhow, we are dealing with a (partially) twisted version of the 
$(0,2)$ supersymmetric theory which we expect to be of a topological type
(see for example \cite{baulieuwest} for the link between
twisting and topological six dimensional QFTs). This suggests that the oscillatory
contribution to the index is $1$ and we will take this point of view. 
Actually, by adapting the results in \cite{gustavsson}
to the relevant quadratic lagrangean of the type considered in \cite{baulieuwest},
it can be checked to happen if $M_4$ is a $T^4$-orbifold
\footnote{Notice that this result should extend to a generic simply connected 
Kaehler $M_4$ by an extension of the methods worked out in \cite{mans,gustavsson}.}.
Therefore, in the subsequent analysis we will
consider the zero-modes contribution.

The contribution to the partition function coming from a single self-dual tensor 
can be analyzed following the results of \cite{witten5,mans}. 
It consists of a $\theta$-function of the lattice of the 
self-dual harmonic three forms.
The $\theta$-function is not completely specified because of the possible 
inequivalent choices of its characteristics {\tiny $\left[\matrix{\alpha\cr\beta}\right]$}.
Notice that the technique developed in \cite{mans} does not in fact single out a particular 
value for the characteristic as an ambiguity in the choice of the relevant holomorphic 
factor. We will find that, in the case at hand, a simple choice 
is automatically made by the requirement of reconstructing the supersymmetric index
that we have reviewed in the previous section.

The relevant $\theta$-function (as calculated in \cite{mans}) is
$$\theta\left[\matrix{\alpha\cr\beta}\right](Z^0|0)
=\sum_k e^{i\pi \left((k+\alpha)Z^0(k+\alpha)+2(k+\alpha)\beta\right)},
$$ where $Z^0$ is a period 
matrix of the relevant six-manifold cohomology that we specify in the following.
Let $\left\{E^{(6)},\tilde E^{(6)}\right\}$ be a symplectic basis 
of harmonic 3-forms on the six-manifold at hand such that, in matrix notation,
$$
\int E^{(6)}E^{(6)}=0,\quad
\int \tilde E^{(6)}E^{(6)}=1,\quad
\int \tilde E^{(6)}\tilde E^{(6)}=0.
$$
We can expand $\tilde E^{(6)}=X^0E^{(6)}+Y^0{}^*E^{(6)}$, where ${}^*$ is the Hodge 
operator. Then $Z^0$ is defined as $Z^0=X^0+iY^0$.

In our case the world volume is in the product form $\Sigma\times M_4$, where $\Sigma=T^2$, 
and therefore, being $M_4$ simply connected, we have
\be
H^3(\Sigma\times M_4)=H^1(\Sigma)\otimes H^2(M_4)
\label{fact}\ee
This means that we can expand $\left\{E^{(6)},\tilde E^{(6)}\right\}$ in terms of a 
symplectic basis
$\left\{[a],[b]\right\}$ for $H_1(\Sigma)$, where
$$
\int_{\Sigma} [a][a]=0,\quad
\int_{\Sigma} [a][b]=1,\quad
\int_{\Sigma} [b][b]=0,
$$
and an orthonormal basis $\{e^{(4)}\}$ for $H_2(M_4)$, i.e. $\int_{M_4} {}^*e^{(4)}e^{(4)}=1$.
In terms of the previous objects we have
$$
E^{(6)}=e^{(4)}\otimes[b]\quad {\rm and}\quad
\tilde E^{(6)}=Qe^{(4)}\otimes[a]
$$
where $Q$ is the intersection matrix 
on $M_4$ given by $Q=\int_{M_4} e^{(4)}e^{(4)}$.
We calculate ${}^*E^{(6)}=-Qe^{(4)}\otimes{}^*[b]$, 
where ${}^*[b]$ is in the two dimensional sense.
By recalling the relation 
$$[a]=-\Omega^{(1)}[b]-\Omega^{(2)}{}^*[b]$$
which holds on every Riemann surface 
\cite{fk}
with period matrix $\Omega=\Omega^{(1)}
+i\Omega^{(2)}$ and the property $Q^2=1$, we get
$$
\tilde E^{(6)}
=
\left(-Q\otimes \Omega^{(1)}\right) E^{(6)} +
\left(1\otimes \Omega^{(2)}\right) {}^*E^{(6)}\, .
$$
Comparing with the general definition we finally read
$$
Z^0=-Q\otimes\Omega^{(1)}+i1\otimes \Omega^{(2)}\, .
$$
In our specific case, since $\Sigma=T^2$ and $\Omega=\tau=\tau^{(1)}+i\tau^{(2)}$ 
is the modulus of the torus, we have simply that 
$$
Z^0=-\tau^{(1)}Q+i\tau^{(2)}1\, .
$$
Now we calculate the relevant $\theta$-function from the zero modes of the
self-dual form in six dimension choosing the zero characteristic candidate
$$\Theta(Z^0)=\theta\left[\matrix{0 \cr 0}\right](Z^0|0)
= \sum_k e^{i\pi k Z^0 k}
$$
Defining $m=k\cdot e^{(4)}$, we calculate $kZ^0k=-\tau^{(1)}(m,m) + i\tau^{(2)}(m,{}^*m)$ 
and we rewrite 
$$
\Theta(Z^0)=
\sum_{m\in\Lambda}q^{\frac{1}{4}(m,*m-m)}{\bar q}^{\frac{1}{4}(m,*m+m)}
$$
where $q=e^{2i\pi\tau}$,
which is equal to (\ref{theta})
as it has been calculated from the reduced dimensional perspectives.

Let us notice that the choice of the null characteristics candidate coincides with 
that of \cite{gustavsson} where it has been shown, in the contest of calculating 
the self-dual tensor partition function on $T^6$, 
to be the only possible contribution leading to a fully modular invariant result.

\subsection{The Little Strings contribution}

To count the full spectrum of the theory a second sector is still lacking.
In fact, the 5-brane theory is completed in the UV by the little string theory
which has BPS saturates strings which eventually have to be kept into account
in the calculation of the complete supersymmetric index.
Even if a full off-shell model for this six-dimensional string theory is not available 
at the moment, we will propose an on-shell simple calculation scheme for the 
supersymmetric index. 

The little string theory model for the world volume theory of the M5-brane 
is built by identifying the boundary states of the membranes ending of the 5-branes with
closed strings configurations whose target space is the 5-brane world volume itself.
These strings are naturally coupled to the self-dual tensor as 
the Poincare' dual of their world-sheet acts as a source for it. This
in fact guarantees the gauge invariance of the effective action for the
5-brane/membrane system.
To calculate the contribution to the supersymmetric index we can proceed as follows.
As it is given by a trace on the string Hilbert space, it corresponds to a one-loop
string path integral. Moreover, as it is usual in these index calculations, 
the semiclassical approximation is exact.
Now, since $M_4$ is simply connected, the only contributions 
to this path integral can arise from string world sheets wrapping the $T^2$
target itself. 
Therefore the configuration space of $n$ of these string world-sheets will be given
by the symmetric product $(M_4)^n/S_n$ whose points parametrize the transverse positions.
Notice that here we are assuming that since there exists only one kind of membranes ending on the 
M5-brane, there is a single type of BPS strings to be counted.
Now, since the supersymmetric index calculated the Euler characteristics
of the configuration space, we claim that the full contribution from these string BPS 
configurations is given by 
$$ 
q^{-\chi_{M_4}/24}
\sum_n q^n \chi\left(M_4^n/S_n\right)=
q^{-\chi_{M_4}/24}
\prod_{n>0}\frac{(1-q^n)^{b_{odd}}}
{(1-q^n)^{b_{even}}}=\eta(q)^{-\chi_{M_4}}$$
where we fixed a global multiplicative factor $q^{-\chi_{M_4}/24}$ because of modularity
requirements and we used well known results from \cite{vw}.

We can compare this result with a natural generalization of 
the construction 
done in \cite{bps5} for toroidal and K3 compactifications to our case.
In fact,
although a covariant quantization scheme does not exists for
six dimensional superstring theory, one can consider its light cone quantization
\footnote{As far as the anomaly cancellation problem 
in a covariant quantization scheme it is concerned, 
it is likely to be solved as indicated in \cite{DVVx}.}
as a temptative good definition. This can be done in our case
since the world volume is in the product form $T^2\times M_4$
by placing the light-cone coordinates along the $T^2$.
Generalizing the construction in \cite{bps5} the twisted superalgebra 
can be obtained from the fermionic zero-modes and the brane supercharges
whose anti-commutation relations 
can be given with a central charge matrix
modeled on the extended intersection matrix
$Z=H\oplus Q$, where 
$H={\tiny\left(\matrix{0 & 1\cr 1 & 0\cr}\right)}$.
Here the two factors corresponds to two-form fluxes and momentum/winding degrees of freedom.
Therefore
we can switch on a set of charges in the lattice
$\Gamma_{b_++1,b_-+1}=\Gamma_{1,1}\oplus\Lambda_{b_+,b_-}$
and, due to the one to one correspondence between charges and fluxes
we can calculate the contribution from the BPS strings. 
In fact, since each BPS saturated string corresponds to a chiral
scalar bosonic mode, these
contribute with a factor of 
$\frac{1}{\eta(\tau)}$ for each possible flux that we can turn on, which is the 
dimension of the above lattice that is
$b_++b_-+2=\chi_{M_4}$.
Therefore we get a total further multiplicative contribution of
$$
\left(\frac{1}{\eta(\tau)}\right)^{\chi_{M_4}}
\,,
$$
which is exactly the same contribution that we have found before.

Multiplying this last factor with that coming from the low energy degrees of freedom
the total supersymmetric index is given by
$$
\frac{\theta_\Lambda}{\eta^\chi}
$$
as given by the dimensionally reduced calculations (\ref{single}).

\section{The multiple M5-brane case}

As far as the multi M5-brane case is concerned, we do not have still a 
direct six-dimensional way to perform
a precise counting similar to the one that we have done for the single M5-brane case
since it is still not known very much about the relevant world-volume theory.
Anyhow, we have some constructive arguments which explain the structure 
of the multi 5-brane index calculation that we review in Section 2.

It is natural to read from these results  
\footnote{See an extended discussion about this point in 
section 4 of \cite{5b}.}
that the moduli space of susy preserving N M5-branes wrapped on a 
supersymmetric Kaehler six manifold is given by 
the space of rank N holomorphic coverings of the manifold itself
plus some (non local) analog of the gauge bundle on them.

This can be checked by the following arguments.
Let us restrict to the case in which the base six manifold admits
a free $S^1$ action and a $S^1$ fibration.
These are the cases where the known formulations for the self-dual two form
\cite{PST,HT} can be formulated without entering further problems
\footnote{That is because in both the cases non zero well defined vector fields 
are required to exist on the manifold to implement the self-duality condition
everywhere.}
and therefore where the $(0,2)$ little string theory admits a better defined 
low energy limit (moreover there are no problems with
anomaly \cite{witten5} since the Euler characteristic
vanishes automatically). 
We compactify further one of the flat transverse directions on a circle that 
we take to be the M-theory eleven coordinate and map the M5-branes to NS5-branes 
in IIA.
Under these circumstances, the $(0,2)$ theory
can be mapped by fiber-wise T-duality to the $(1,1)$ little string theory
which represents the decoupled limit of the NS5-branes of IIB.
By the S self-duality of type IIB we map this system to
an equivalent system of D5-branes in type IIB.
In fact, D5-brane bound states are described, in the low energy approximation,
by susy-preserving configurations of the dimensional reduction of the 
${\cal N}=1$ U(N) ${\rm SYM}_{10}$ to the susy-cycle ${\cal C}$ on which they are wrapped on.
In particular this means that the spectrum of the transverse bosonic fields define
an holomorphic covering of rank $N$ in the total space of the normal bundle 
of ${\cal C}$ in the ambient ten dimensional manifold (which is the ambient manifold itself).
We naturally expect that the UV completition of the S-dual 
$(1,1)$ theory still contains this BPS geometric moduli space naturally
and therefore also its fiber-wise T-dual $(0,2)$ little string theory
that we started from.

From this perspective one could try to speculate about the possible structure of the 
analog of the gauge bundle structure. In the case of a single M5-brane, it is given by
a higher dimensional analog of the Abelian gauge bundle structure realized
within the 1-form valued Cech cohomology
\footnote{This means that the two form potential undergoes a gauge like
transformation from patch to patch and that the patching forms satisfy
certain consistency conditions.}
. As it seems by now, there does not exist any 
straightforward non-Abelian analog for it (at least in a local framework generalization
\footnote{Notice the strong resemblance with the no-go theorem in \cite{BHS} about 
local interactions between chiral forms.}).
A possible way out is given anyhow by the pullback procedure once the covering map
from the base cycle to the multi 5-brane world-volume is given.
Notice that the pull back map fails exactly at the branching locus of the covering,
where in the typical D-brane cases the full non-Abelian structure of the underlying 
vector bundle becomes crucial, and therefore we can't refer to any well 
defined local geometrical structure on the base cycle to be understood to be the pull back 
of a gerbe on the covering cycle. We conclude therefore that a non-local completition of the 
pull back map should be accounted for by the UV non-local structure of the theory.

In the case in which the world volume manifold is in the product form
$T^2\times M_4$, with $M_4$ Kaehler and simply connected, the formula
(\ref{eg}) applies to the supersymmetric index. 
The $T^2$ holomorphic self-covering is unbranched and
 the little strings
contribution is explicitly exposed and we interpret it also to encode
the above mentioned effect. In fact it enters in the form 
$\left(\frac{1}{\eta}\right)^{\chi_{\Sigma_a}}$, where $\Sigma_a$ is a rank $a$ holomorphic 
covering of $M_4$
(spectral manifold). The explicit dependence on the branching locus $B_a$ appears
just because 
$$\chi_{\Sigma_a}=a\cdot\chi_{M_4}-\chi_{B_a}\,.$$ 

We can compare positively all these arguments with the approach developed in \cite{agmh}
where BPS stringy representations of the $(0,2)$ algebra are obtained 
and the natural role of string degrees of freedom in encoding the very
structure of the interacting M5-brane world-volume theory is enlighten.

Let us notice moreover also the following consequence of the construction 
given in the previous section, which appears 
once we compare it with the geometrical moduli space
which we calculated in \cite{5b}. Once we consider this moduli space from the
point of view of the little string theory BPS states, we find these strings
to wrap along the $T^2$ and join/split in points located 
at the branching locus of the $M_4$ covering.
Notice that this behavior is typical of the matrix strings \cite{mst,MST} and 
is coherent with the possibility of generalizing the approach in \cite{5dbh}
to our case. Unfortunately, the problem of the formulation of Matrix Theory
on curved manifold is still far from being understood, but it turns out
that these two similar structures likely correspond each other
under the M-theory electric/magnetic duality
which exchanges membranes and 5-branes.

As a consequence of the above observations, it is therefore natural to conjecture that 
the little string theory supersymmetric index (and moreover also a wider set of correlation 
functions) can be built from $\theta$-functions of the period matrix of the covering 
manifolds by combining them both via their characteristics 
(see the structure of (\ref{eg}) where the shifted lattice corresponds to non zero characteristic
for the relevant $\theta$-function)
and via the covering structure itself
as far as the UV completition with the little string contributions
is concerned. 
In particular one expects that the above ratio works 
for the calculation of correlators of self-dual strings in the form of 
surface operators in the low energy approximation.

We expect then that, for any given rank $N$ covering ${\cal C}$ of the world-volume
manifold on which the $N$ 5-branes are wrapped, the contribution to the 
supersymmetric index is of the form
\be
\frac{\theta\left[\matrix{\alpha \cr \beta}\right](Z^0_{\cal C}|0)}{\N(Z^0_{\cal C})}
\label{banf}\ee
where 
$\N(Z^0_{\cal C})$ is a modular form built from the periods of the covering
six manifold which
generalizes the little strings contribution to the generic case
while the non-zero characteristic ${\tiny \left[\matrix{\alpha \cr \beta}\right]}$
encodes the twist induced by 
inequivalent spin structures \cite{ww,wm,hsix}
analogous to the one which appears in (\ref{eg}).
These structures and the possible rank $N$ holomorphic coverings has then to be summed up.

We expect also that a structure analogous to (\ref{banf}) applies also to the set of amplitudes
relative to 
a proper supersymmetric version of the surface operators considered in 
\cite{surf,mans} 
where a non zero argument of the $\theta$-function 
encodes the surface periods.

\section{Conclusions and open questions}

In this note we have proposed a six dimensional framework 
for the evaluation of the supersymmetric index of the M-theory
five-brane which correctly reproduces the results in \cite{5b}
and
proved the explicit need of taking into account the full spectrum 
of the little string theory to reproduce a precise counting
in the form of BPS saturated string states.

It would be very interesting of course to check how the methods that we developed 
in this note extend to other possible M5-brane geometries
and to continue the analysis of the Little String theory  
to try to better understand the several unclear points which are left over.
Possible interesting configurations 
which naturally generalize the one we have studied here
could be M-theory
on $R^3\times Y$, with $Y=CY_4, K3\times K3$ and the 5-branes
wrapped on a six holomorphic cycle in $Y$.

The question related to how one should exactly build (at least a set of)
correlation functions in little string theory by using these $\theta$-function 
building blocks remains open and needs a much more accurate analysis.
For an example in a close-by perspective see \cite{hsix}.
In particular one expects that the above ratio works 
for the calculation of correlators of self-dual strings in the form of 
surface operators.

Another important issue which is raised by taking seriously the Little String Theory
hypothesis is the following. Since it is a superstring theory in 6 dimensions,
it is non critical and this means that one should 
\footnote{As an alternative, one might try to add to the theory a non geometrical sector
to reach the Virasoro central charge saturation, but these degrees of freedom
should not contribute both to the low energy effective theory
and neither to the supersymmetric index.}
take into account also 
the Liouville sector which does not decouple from the $\sigma$-model.
Notwithstanding it has been extensively studied,
the theory of the Liouville (super-)field still lacks a full solution and this
problem, which in critical perturbative superstring theory was token for avoided,
seems to come back into the game again.
Let us notice here just the following coincidence.
The near horizon geometry of the M5-brane in eleven dimensional super-gravity
is $AdS_7\times S^4$. Suppose we study the coupled theory of the 5-brane
as a membrane theory on this fixed background. Then the longitudinal 
radial membrane field spans the radial direction in the $AdS_7$.
The boundary value of this field will presumably play a crucial role 
in a decoupling procedure and it should be linked to the Liouville field
which appears in the six dimensional string theory in a way similar to
the original point of view of \cite{polyakov}.
An apparently unrelated question concerns the claim that the
low energy limit of little string theory does not contain 
gravitational degrees of freedom. This issue could also be related
to the specific nature of the full six dimensional non-critical
string theory.

\vspace{.5 cm}
{\bf Acknowledgments}
I would like to thank M.~Bertolini, A.M.~Boyarsky, 
J.F.~Morales, A.~Tomasiello and M.~Trigiante for interesting discussions.
Work supported by the European Commission RTN programme
HPRN-CT-2000-00131.

\end{document}